\documentclass[twoside]{article}
\usepackage{fleqn,espcrc2}

\usepackage{epsfig}

\newcommand{\beeq}{\begin{equation}}
\newcommand{\eneq}{\end{equation}}

\newcommand{\AmS}{{\protect\the\textfont2
  A\kern-.1667em\lower.5ex\hbox{M}\kern-.125emS}}

\title{Current Status of the Numerical Simulations of d=3 SU(2) Lattice Gauge Theory
in the Dual Formulation. }

\author{N.D. Hari Dass\address[IMSC]{Institute of
Mathematical Sciences, 
        C.I.T Campus,\\ 
        Taramani, Chennai 600 113, India}%
\thanks{{\bf email:dass@imsc.ernet.in}}
        and
        Dong-Shin Shin\addressmark[IMSC] 
\thanks{{\bf email:shin@imsc.ernet.in}}}
       
\begin{document}

\begin{abstract}
We have continued our systematic investigations of the numerical simulations
of lattice gauge theories in the dual formulation. These include: i) a more
practical implementation of the quasi-local updating technique, ii) a thorough
investigation of the sign problem, iii) issues related to the ergodicity of
the various update algorithms, iv) a novel way of measuring conventional
observables like plaquette in the dual formalism and v) investigations 
of thermalisation. While the dual formulation holds out a lot of promises
in principle, there are still some ways to go before it can be made into
an attractive alternative lattice formulation.
\vspace{1pc}
\end{abstract}

\maketitle

\section{INTRODUCTION}
We are reporting on our continuing investigations of $SU(2)$ lattice gauge 
theory in the dual formulation. A brief introduction to the formalism
as well as to the techniques that have been developed for numerical
simulations was presented at LATTICE'99 \cite {lat99}. As was stated there, 
the partition function of the conventional LGT can be converted,
upon using the techniques of character expansion and group integrations,
into the dual form \cite{anishetty} (for another perspective
on this issue see \cite{pfeiffer}):
\beeq
Z_{d} =\sum_{\{j\}}\prod (2j+1)C_{j_a}(\beta)
\prod_{i=1}^{5}\left\{\matrix{a_i&b_i&c_i\cr d_i&e_i&f_i\cr }\right\}
\eneq
This partition sum is defined over the dual lattice where $\{j_a\}$ live over
the links and $\{j_b\}$ over the diagonals to the plaquettes. The convention for
the diagonals is that they connect the vertices of the odd sublattice. We have
collectively designated $\{j_a\},\{j_b\}$ by $\{j\}$. Each cube of the dual lattice
is seen to be spanned by 5 tetrahedra of which one is spanned entirely by $\{j_b\}$
while four are spanned by three $\{j_a\}$ and three $\{j_b\}$. These
features are illustrated in fig.1 which exhibits a two-cube cell on the
dual lattice. The b-links are shown by dashed lines while the a-links
are shown by solid lines.
Each tetrahedron
carries a weight \mbox{factor which is the $SU(2)$ 6-j symbol $\left\{\matrix{a&b&c\cr
d&e&f\cr}\right\}$.} \mbox{Periodic b.c for the original lattice is} 
crucial for this construction.

One immediate advantage of the dual approach is that variables are
now discrete and for all practical purposes can be represented by
short integers. Other advantages of this method have already been
commented upon in \cite{lat99}.
\subsection{Conventional Observables}
Usually when one goes to the dual description, observables in the
original
formulation become very complicated in the dual picture and vice
versa (for a discussion of the Wilson loop in the dual picture see
\cite{pfeiffer}). It is therefore curious to see that, in the present context, a
whole class of conventional observables are just as easily measured
in both the formulations. In the conventional formulation, 
the expectation value of the normalized plaquette is given by
\beeq
<P> = {1\over N_p}{\partial \log Z(\beta)\over \partial \beta}
\eneq
where $N_p$ is the number of plaquettes in the lattice. In the dual
picture it can be easily verified that this expectation value is
given by
\beeq
<P> = <{d \log C_j(\beta)\over d\beta}>
\eneq
where the average on the r.h.s is over the distribution of $j$-values
on the a-links of the dual lattice. This formula can be generalized to the
case of any function of the plaquette variable and more importantly
to the two-point correlation functions of plaquettes or of functions
of plaquette variables. This allows for a direct comparison between
the dual and the conventional pictures.

\section{UPDATES}
In \cite{lat99} we introduced two types of updating algorithms:
{\bf local} and {\bf quasi-local}. It was already pointed out there
that the {\bf local} updates are not ergodic as they are unable to
change integer spins to non-integer spins and vice versa. We
illustrate the {\bf local} update algorithm with the help of fig.1.
\begin{figure}[htb]
\begin{center}
\mbox{\epsfig{file=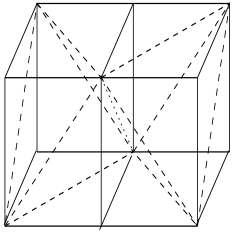,width=3truecm,height=3truecm,angle=-90}}
\caption{Illustrating the local updates. }
\label{Fig 1.}
\end{center}
\end{figure}

The b-link shown as a dotted line is contained in 6 tetrahedra
(two of them containing only b-links); it is also contained in 6
triangles again two of which contain only b-links. For each triangle
containing this link, one forms $j^i_{max}= j^i_1+j^i_2$ and 
$j^i_{min} = |j^i_1-j^i_2|$ where $j^i_{1,2}$ are the remaining two
spins in the $i$-th triangle containing the link to be updated. One then
forms $j_{max} = min_i\{j^i_{max}\}$ and $j_{min} = max_i\{j^i_{min}\}$
and updates the link to a value in the range $(j_{min},j_{max})$.
Clearly this does not change half-integers to integers and vice
versa and the move is not ergodic by itself.

In addition to the problem of ergodicity, there are two additional
problems with this method. Every now and then one may face a
situation where $j_{min}=j_{max}$ in which case the move can not
change the value of the link. Our code keeps track of this possibility
through a counter {\bf irange} and whenever this happens one skips
the particular site and moves on to the next one. The other problem
has to do with the fact that 6j's occasionally vanish even when all
the triangle relations hold. Very rarely it so happens that every new
value of the link is such that the product of 6j's over the
tetrahedra (four in the case of local updates of a-links and six in
the
case of b-links) vanishes and the new configuration is to be
eliminated because it does not contribute to either the partition sum
or to any of the expectation values. This is kept track of by the
counter {\bf ispecial}.

\subsection{Quasi-local Updates}
As already discussed in \cite{lat99}, if we were to change a link from
half-integer (henceforth referred to as fermionic) to integer (bosonic)
or vice versa, in general the change proliferates throughout the
lattice as at least two links have to be $Z_2$-flipped in every
triangle. What one needs are flippings involving the smallest cluster
of links.Two categories of such {\bf quasi-local} updates were
identified in \cite{lat99}. At an even site of the lattice, the update
algorithm called {\bf globeven} involves the interior links of a
cluster of 8 tetrahedra as shown in fig.2.
\begin{figure}[htb]
\begin{center}
\mbox{\epsfig{file=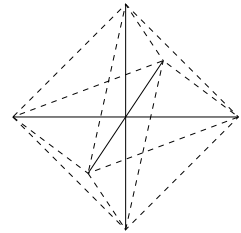,width=3truecm,height=3truecm,angle=-90}}
\caption{Quasi-local updates at even sites.}
\label{Fig 2.}
\end{center}
\end{figure}
Likewise, at odd sites the corresponding update is called {\bf globodd}
and it involves the 6 a-links and 12 b-links in the interior of the
8-cube cluster involving 32 tetrahedra as shown in fig.3.
\begin{figure}[htb]
\begin{center}
\mbox{\epsfig{file=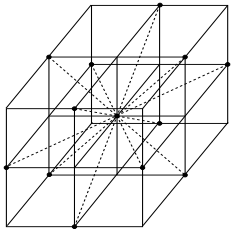,width=3truecm,height=3truecm,angle=-90}}
\caption{Quasi-local updates at odd sites.}
\label{Fig 3.}
\end{center}
\end{figure}
In both these schemes of quasi-local updates, one keeps all the links
on the outer surface fixed and $Z_2$-flips all the interior links. The
changes should be so made as to satisfy all the triangle inequalities.
In \cite{lat99}, the basic idea of achieving this goal was briefly
mentioned. This is done using variables originally introduced by
Bargman\cite{barg} and developed by \cite{gadiyar}. This will be discussed
in detail in \cite{hari2K}. While this method based on the
K(agome)-variables is the systematic approach to the quasi-local
updates, it is pretty involved and time consuming. A simplified
algorithm was proposed in \cite{lat99} which consisted of simply
shifting all spin values on the interior links by ${1\over 2}$
which can be seen to preserve all the triangle inequalities. But
this can result in substantial changes in the Boltzman weight
resulting in abrupt changes. In the current investigations we modified
this by following the fixed $Z_2$-flip by randomly selecting a set
of configurations satisfying the triangle inequalities and then
performing heatbath or metropolis in the subspace of these
configurations.
\section{THE SIGN PROBLEM }
It was already emphasised in \cite{lat99} that one of the main
drawbacks of the dual formulation is the lack of positivity of the
Boltzman weight which arises out of the 6j's not being of a particular
sign. There we had proposed to use the trick of generating important
samples with probabilities that are the modulus of the weights
occurring in eqn (1) but by absorbing the sign of the configuration
into the observables. It was soon pointed out to us by Wosiek
\cite{wosiek} that no matter how small the local probability of a
wrong sign, in the thermodynamic limit there will be an almost equal
fraction of configurations with the two signs. This meant that the
average value of the sign is very close to zero making the procedure
questionable. One way out is to regroup the terms in eqn(1) so that 
for each group the weight is positive. But it is not clear how to do
this. To make some progress we decided to adopt another strategy based
on the fact that sign changes in the configurations come about through
completely local sources. To see this, let us look at the expression
for the average of a generic observable O:
\beeq
\langle O \rangle_p = {\langle O\epsilon \rangle_{\tilde p}\over \langle \epsilon \rangle_{\tilde p}}
\eneq
where $\{\tilde p_i={|p_i|\over P}\}$ with $P=\sum_i |p_i|$ and
$p_i$ are the weights occurring in eqn(1) and are not necessarily
positive. Further, $\epsilon_i$ is the sign of $p_i$. Now if it so happens 
that when thermal equilibrium is reached
the average value of $O$ (in the usual sense) is insensitive to the
sign of the configuration in the sense that the average over positive
sign configurations is very close to the average over negative sign
configurations, the numerator in eqn(4) is also proportional to
$\langle \epsilon \rangle_{\tilde p}$ and we can simply ignore the
sign problem.

We wish to argue that this is very likely to happen for relatively local
observables and also for the correlation functions of relatively local
observables. 
The reason for this is the fact that the source for the change of sign of the 
configuration is ultralocal. Combining this with the expectation that both
signs become equally likely as the system size increases, one is led to surmise that
the configurational average of reasonably local observables as well as the
correlation functions of reasonably local observables are insensitive to
the sign of the configuration. So far most of our runs have borne out this
expectation. We found that whenever there was a reasonable fraction of 
configurations with each sign, the averages of inner and outer tetrahedra 
were indeed insensitive to the sign in the limit of thermalisation
(to be more precise, in the limit of nearly stationary expectation values).
We are investigating this with other classes of observables. We also propose
to make a systematic study of the sign problem by studying the way the signs
fluctuate locally as well by studying their correlations over the lattice.

\section{METROPOLIS AND HEATBATH}
For the local updates both the metropolis and heatbath methods work in  
fairly standard ways because the allowed range of values for the new link
are very small. But for the quasi-local moves the situation is quite
different. Even in the case of the quasi-local updates at even sites, we
are dealing with the space of allowed spin values on six links. Truncating
the maximum value of spins at ${9\over 2}$ as we did, we have $10^6$ values
to check for triangle inequalities and subsequently picking a new element
from among the list. This is a hopeless task. Even using the K-variable
method\cite{hari2K} is time consuming. So what we did was to apply the fixed $Z_2$-flip
and select the first a-link randomly from its allowed possibilities, the second a-link
randomly from its allowed values (which depends on the random choice made for the first
a-link) and so on. For metropolis algorithm, after selecting a sextuplet of values,
the new set was accepted or rejected according to the usual metropolis rule.
For the
heatbath method we likewise selected a certain number of sextuplets before the $Z_2$-flip 
and an equal number after the flip. The heatbath algorithm was then applied to this
manageable subset. At present we do not have a more rigorous argument to justify 
our heatbath method. For the globodd update the problems are even more severe and we
have again adopted the same pragmatic approach.
\subsection{INITIAL CONFIGURATIONS}
As initial configurations we chose from among three classes: i) cold  start - here
we set all the a-links and b-links to $j=1$. This configuration clearly satisfies
all the triangle inequalities. ii) extreme cold start - here all a-links and b-links
were set equal to $j =0$. It is worth noting that the extreme cold start configuration
is a fixed point of the local update algorithms. iii) hot start - here a-links and b-links are
randomly chosen in a way that all triangle inequalities are satisfied. A practical way
of reaching such configurations is to first start with a cold start configuration
and go on choosing links randomly in their allowed ranges. After sweeping the lattice 
sufficiently many times one gets to a hot start configuration.
\section{RESULTS}
We carried out our simulations on a $8^3$ lattice at various $\beta$-values.
We simulated 
with the absolute values of the weights given by eqn(1) and monitored
the fluctuations in the sign of the configurations. In a few runs we noticed that the sign
changed very few times. In those cases, the average values of the inner (outer) tetrahedra
for the positive sign and negative sign configurations did not approach the same values even after
many iterations. But for these cases 
$\langle \epsilon \rangle_{\tilde p}$ 
is not close to zero and one can safely use eqn (4). On the
other hand for a majority of runs, the sign fluctuated quite a bit and even when 
$\langle \epsilon \rangle_{\tilde p} \simeq 0.5 $, the average values for the two cases
were very close to each other bearing out the surmise made earlier. This was so for all
the classes of update algorithms, initial states and metropolis/heatbath. We plan to study
this issue for a larger class of observables.

In our simulations using only the local updates, we found that while the results of
metropolis and heatbath converged for given initial states, the results did not converge
for trials with different initial states. We have interpreted this as due to the
non-ergodic nature of the local updates. The results for the quasi-local updates were not
so clear cut. We also ran the code with the updating algorithms being chosen randomly;
at the first instance we have run the code such that all updating algorithms are
equally likely. Here we find that while the runs for the cold-start and hot-start converged,
the results for metropolis and heat bath are widely divergent. We have not really understood
the reason for this.

\section{CONCLUSION}
Clearly a lot more work needs to be done before establishing the dual framework as
a viable alternative for numerical simulations of Gauge and Spin systems. The most
pressing problem is to gain a satisfactory understanding and control of the sign 
problem. While our surmise in this direction is well borne out by the initial runs,
we are yet to show its correctness in a controlled manner. One interesting possibility
is to study the model defined by the absolute value of weights in its own right
and we plan to undertake that study shortly. But we still have to understand the
lack of convergence between the different runs before we can take any of the conclusions
seriously. To this end, we are going to investigate the efficacy of our metropolis
and heat bath methods by studying them in contexts where we can get results by
deterministic as opposed to stochastic methods. 

Another line of attack we are
contemplating is to look for simple models that are analytically tractable but
which nevertheless have all the essential ingredients of the dual formulation 
studied here. Indeed, such models were encountered in the study of analytic variational
methods for euclidean lattice gauge theories \cite{Hari84} and in the analytic evaluation of the
partition function of the unit hypercube for $d=4$ $SU(2)$ lattice gauge theories \cite{hyper}.
We are hoping that the insights so gained will enable us to complete the
investigations reported here.

There are a large class of very important problems that belong to the same calculational
scheme as what we have attempted here. The most notable are: the $d=4$ version of these results
as obtained by Halliday and Suranyi\cite{halliday}, the class of models considered by 
Ooguri \cite{ooguri} and their non-topological descendants etc.Language and techniques very
similar to the ones used in the dual picture are also becoming increasingly popular
in treatments of quantum gravity \cite{baez}. Replacing the 6j's by q-deformed 6j's
would immediately relate these developments to the well known Turaev-Viro models and their
generalizations. Clearly it is of great importance to make further progress in the numerical
simulations of theories in the dual formulation.

\section{ACKNOWLEDGEMENTS}
NDH wishes to thank Dick Haymaker, J. Wosiek, Erhard Seiler, Peter Weisz, 
Srinath Cheluvaraja, A. Hart, N. Kawamoto and Maarten Golterman for many
interesting discussions. We are thankful to all those who visited our posters at LATTICE2000 for the
lively discussions.

\end{document}